\documentclass[epj,twocolumn]{webofc}
\usepackage[varg]{txfonts}   % Web of Conferences font
\doi{}
\usepackage{subfig}
\usepackage{comment}
\usepackage{color}
\usepackage{subfig}

\usepackage{hyperref}
\hypersetup{
  pdftitle={},%
  pdfauthor={},%
  pdfsubject={},%
  pdfkeywords={},%
  pdfstartview={},%
  bookmarksopen=true, breaklinks=true, debug=true, %
  colorlinks=true, linkcolor=blue, citecolor=blue, urlcolor=blue
}

\newcommand{\cf}[1]{{Fig.~\ref{#1}}}

\newcommand{\glabcms}{\gamma^{\rm lab}_{\rm c.m.s.}}
\newcommand{\blabcms}{\beta^{\rm lab}_{\rm c.m.s.}}
\newcommand{\dylabcms}{\Delta y^{\rm lab}_{\rm c.m.s.}}

\newcommand{\ie}{{\it i.e.}}
\newcommand{\eg}{{\it e.g.}}
\newcommand{\etal}{{\it et al.}}

\woctitle{TRANSVERSITY 2014} 
\begin{document}
\title{Spin physics and TMD studies at A Fixed-Target ExpeRiment at the LHC \\(AFTER@LHC)}
%
% subtitle is optional
%
%\subtitle{}

\author{
J.P. Lansberg\inst{1} \thanks{\email{Jean-Philippe.Lansberg@in2p3.fr}}
\and
M. Anselmino\inst{2}
\and
R.~Arnaldi\inst{2}
\and
S.J.~Brodsky\inst{3}
\and
V.~Chambert\inst{1}
\and
W.~den Dunnen\inst{4}
\and
J.P.~Didelez\inst{1}
\and
B.~Genolini\inst{1}
\and
E.G.~Ferreiro\inst{5}
\and
F.~Fleuret\inst{6}
\and
Y.~Gao\inst{7}
\and
C.~Hadjidakis\inst{1}
\and
I.~Hrvinacova\inst{1}
\and
C.~Lorc\'e\inst{3,8,1}
\and
L. Massacrier\inst{9,10,1}
\and
R.~Mikkelsen\inst{11}
\and
C.~Pisano\inst{12}
\and
A.~Rakotozafindrabe\inst{13}
\and
P.~Rosier\inst{1}
\and
I.~Schienbein\inst{14}
\and
M.~Schlegel\inst{4}
\and
E.~Scomparin\inst{2}
\and
B.~Trzeciak\inst{15}
\and
U.I.~Uggerh\o j\inst{11}
\and
R.~Ulrich\inst{16}
\and
Z.~Yang\inst{7}
}

\institute{
IPNO, Universit\'e Paris-Sud, CNRS/IN2P3, F-91406, Orsay, France
\and
Dip. di Fisica and INFN Sez. Torino, Via P. Giuria 1, I-10125, Torino, Italy      
\and
SLAC National Accelerator Lab., Theoretical Physics, Stanford University, Menlo Park, CA 94025, USA  
\and
Institute for Theoretical Physics, Universit\"at T\"ubingen, Auf der Morgenstelle 14, D-72076 T\"ubingen, Germany
\and
Departamento de F{\'\i}sica de Part{\'\i}culas, Universidade de Santiago de C., 15782 Santiago de C., Spain
\and
Laboratoire Leprince Ringuet, \'Ecole Polytechnique, CNRS/IN2P3,  91128 Palaiseau, France
\and
Center for High Energy Physics, Department of Engineering Physics, Tsinghua University, Beijing, China
\and
IFPA, AGO Dept., Universit\'e de Li\`ege, Sart-Tilman, 4000 Li\`ege, Belgium
\and
SUBATECH, Ecole des Mines de Nantes, Universit\'e de Nantes, CNRS-IN2P3, Nantes, France
\and
LAL, Université Paris-Sud, CNRS/IN2P3, Orsay, France
\and
Dept. of Physics and Astronomy, University of Aarhus, Denmark 
\and
Nikhef and Dept. of Physics \& Astronomy, VU University Amsterdam, NL-1081 HV Amsterdam, The Netherlands
\and
IRFU/SPhN,CEA Saclay, 91191 Gif-sur-Yvette Cedex, France
\and
LPSC, Universit\'e Joseph Fourier, CNRS/IN2P3/INPG, F-38026 Grenoble, France
\and
FNSPE, Czech Technical U., Prague, Czech Republic
\and
 Institut f\"ur Kernphysik, Karlsruhe Institute of Technology (KIT), 76021 Karlsruhe, Germany
}

\abstract{%
We report on the opportunities for spin physics and Transverse-Momentum Dependent distribution (TMD) 
studies at a future multi-purpose fixed-target experiment using the 
proton or lead ion LHC beams extracted by a bent crystal. The LHC multi-TeV beams allow for the most energetic 
fixed-target experiments ever performed, opening new domains of particle and nuclear physics and complementing 
that of collider physics, in particular that of RHIC and the EIC projects. The luminosity achievable with 
AFTER@LHC using typical targets would surpass that of RHIC by more that 3 orders of magnitude in a similar energy region.
In unpolarised proton-proton collisions, AFTER@LHC allows for measurements of TMDs such as the Boer-Mulders quark 
distributions, the distribution of unpolarised and linearly polarised gluons in unpolarised protons. Using the 
polarisation of hydrogen and nuclear targets, one can measure transverse single-spin asymmetries of quark and 
gluon sensitive probes, such as, respectively, Drell-Yan pair and quarkonium production. The fixed-target mode has the 
advantage to allow for measurements in the target-rapidity region, namely at large $x^\uparrow$ in the polarised nucleon. Overall, 
this allows for an ambitious spin program which we outline here.
}
\maketitle
\section{Introduction}
\label{intro}

More than ten years ago, RHIC opened a new era in the study of spin physics at relativistic energies in
being the first collider of polarised protons. Thanks to the polarisation of both beams, double-spin 
asymmetries could be measured (see~\eg~\cite{Adler:2004ps,Abelev:2006uq}) and, thanks to 
its high center-of-mass energy --up to 500 GeV--, 
the measurements of spin asymmetries in weak boson production could be performed~\cite{Adare:2010xa,Aggarwal:2010vc}.

Unfortunately, neither the Tevatron nor the LHC were designed with the possibility of colliding
polarised  protons. Nevertheless, it has recently been emphasised that a class of spin-dependent partonic 
distributions can be studied even in the absence of polarised proton. In fact, the polarised particle is,
in this case, the parton in an unpolarised nucleon and a correlation effect arises for nonzero partonic 
transverse momenta. In the quark case, these quantities are named Boer-Mulders distributions~\cite{Boer:1997nt}. 
In principle, the LHC machine can thus also be used to perform spin-related
measurements.

Much more can however be done~\cite{Brodsky:2012vg} if the multi-TeV proton LHC beams are extracted and sent to a fixed 
target, the latter being polarised or not. In the former case, one can study a number of target (transverse) spin
asymmetries, also called single transverse spin asymmetries (STSA). In the latter case, since the 
typical conditions of a fixed-target experiment allow for rather low transverse-momentum measurements,
one can perform a number of studies of Boer-Mulders function for the quark sector or of the distribution
of polarised gluons in unpolarised nucleons.

In this context, it is useful to recall the critical advantages of a fixed-target experiment 
compared to a collider one, \ie~
\begin{itemize} \itemsep-4pt
\item[-] extremely high luminosities thanks to the high density of the target; 
\item[-] absence of geometrical constraints to access the far backward region in the c.m.s.;
\item[-] unlimited versatility of the target species; and
\item[-] same energy for proton-proton, proton-deuteron and proton-nucleus collisions.
\end{itemize}

These first two advantages are particularly relevant for the topics 
to be discussed here and discussed in \cite{Lorce:2012rn,Rakotozafindrabe:2013au}, whereas 
the latter two are more relevant for heavy-ion physics previously discussed 
in~\cite{Lansberg:2012kf,Rakotozafindrabe:2012ei,Lansberg:2013wpx}.

\section{Beam extraction and target polarisation}

The extraction of beams by using the technique of  bent-crystal channelling 
offers an ideal and cost-effective way to obtain a clean and very 
collimated beam even at TeV energies. This exhibits the asset of not altering  the LHC beam 
performances~\cite{Uggerhoj:2005xz,LUA9}. 
A "smart collimator" solution will be tested on the 7-TeV LHC beam by 
the CERN LUA9 collaboration after the current long shutdown (LS1)~\cite{LHCC107}; a minimal setup 
that includes a horizontal and a vertical piezoelectric goniometre with the associated crystal
have already been installed in the LHC beampipe in IR7~\cite{LMC-173}. 
Another proposal, to be further investigated, is to "replace" the kicker-modules in LHC section IR6 
by a bent crystal~\cite{Uggerhoj:2005xz}.

In terms of kinematics, 7 TeV protons colliding 
on fixed targets release a center-of-mass energy close to 115~GeV ($\sqrt{2E_p m_N}$).
The extraction has also been tested for heavy-ion beams, for instance at SPS 
by  the CERN UA9 collaboration~\cite{Scandale:2011zz}. The 2.76 TeV LHC lead beam
would for instance allow one to study heavy-ion collisions at a center-of-mass energy per 
nucleon-nucleon collision close to 72 GeV, exactly half way between RHIC and SPS experiments.

The extraction procedure is as follows: one would position a bent crystal 
in the halo of the beam such that a few protons (or lead) per bunch
per pass are channelled in the crystal lattice. These are consequently deviated by a couple of
mrad w.r.t. to the axis of the beam. A significant fraction of the beam loss can then 
be extracted likewise, with an intensity of $5\times 10^8$ $p^+$s$^{-1}$. This corresponds 
to an average extraction per bunch per revolution of mini-bunches
of about 15 $p^+$ each 25 ns.

Past experiments (see \eg~\cite{Baur00}) have shown that the  degradation of the crystal is at the level of $6\%$ per 
$10^{20}$ particles/cm$^2$. Such an integrated intensity is equivalent to a year of operation, 
for realistic impact parameters and realistic beam sizes at the location of the crystal. 
After a year, the crystal has to be moved by less than a 
millimetre such that the beam halo hits the crystal on an intact spot. This procedure can
be repeated without specific constraints.

Despite the outstanding luminosities which can be obtained, the intensity of the extracted beam 
is not extremely large. In particular, it  does not constrain the choice of the target-polarisation technique.
With such a highly energetic beam, one expects a minimum ionisation and a low heating of the target. 
The expected heating power due to this extracted beam is on the order of 50 $\mu\hbox{W}$ for a typical 
1cm thick target. Temperatures as low as 50 mK  can thus be maintained in the target. 
In the spin-frozen mode, relaxation times can last as long as one month.
The damages on the target would typically arise after an irradiation of $10^{15} p^+ \hbox{cm}^{-2}$~\cite{meyer}; 
this corresponds to 1 month of exposition in this case.

Yet, one cannot ignore the major constraint set by the available space in the underground LHC complex. 
This most likely restricts the choice  to polarisation by continuous {\it Dynamic Nuclear Polarisation} DNP 
or to a HD target~\cite{didelez}. Both take less space than the frozen-spin  machinery. 
The project AFTER@LHC is a strong motivation to revisit  the necessary technology~\cite{solem}.
CERN has a long tradition of DNP for a number of materials such as NH$_3$ and Li$_6$D~\cite{berlin}.  
Experts of DNP can still be found worldwide, while 
HD target makers are more rare, \eg~one at TJNAF (USA) and the other at RCNP (Japan)~\cite{kohri}.

\begin{figure}[!hbt]
\includegraphics[height=\columnwidth,angle=-90]{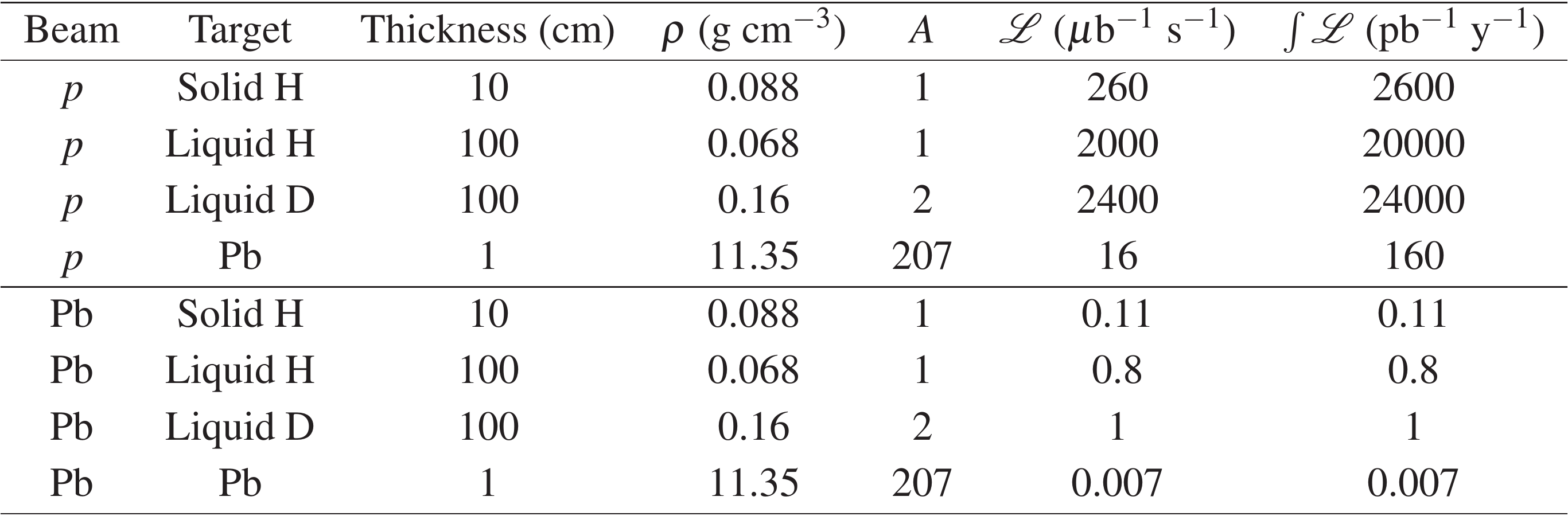}
\caption{Instantaneous and yearly luminosities obtained for targets of various thickness 
with an extracted beam of (a)
$5 \times 10^8$ p$^+$/s with a momentum of 7 TeV and (b) $2 \times 10^5$ Pb/s with a momentum per nucleon of 2.76 TeV.}
\label{tab:lumi}
\end{figure}

The instantaneous and yearly (over 10$^7$~s) luminosities reachable
with the proton beam on targets of various thickness are gathered in table~\ref{tab:lumi}. 
Note that 1m long targets of liquid hydrogen or deuterium
 give luminosities close to 20 femtobarn$^{-1}$, as large as the luminosities collected at the LHC at 7 and 8 TeV.
Table~\ref{tab:lumi} also gathers the corresponding values for the 
Pb run of 10$^6$~s. 

\section{Short selection of highlight studies not related to spin.}

Before discussing measurements pertaining to spin and TMD physics, it is instructive to recall what can, in principle, 
be done to learn more about the spin-independent inner structure of nucleons and nuclei. This section  only presents 
an introductory selection of studies relevant for the discussions of the next sections. A more 
complete survey of the physics opportunities with AFTER@LHC can be found in~\cite{Brodsky:2012vg}.

Given the possibility of studying a number of colliding systems, such as proton--proton,
 proton--deuteron, lead--proton, lead--nucleus, proton--polarised-nucleon and lead--polarised-nucleon, 
we believe that AFTER@LHC should be designed as a multi-purpose detector. As such, 
it would {\it de facto} become  a heavy-flavour, quarkonium and prompt-photon 
{\it observatory}~\cite{Brodsky:2012vg,Lansberg:2012kf} in 
$pp$ and $pA$ collisions given the outstanding expected luminosity combined to an access towards low $P_T$.
In turn, thanks also to the complementary forthcoming LHC results, it is sound to expect that 
the production mechanisms of quarkonia~\cite{review} would eventually be constrained thanks to the large quarkonium yields
 and precise measurements of their correlations.

With 7 TeV protons,  the boost between c.m.s. and the lab system is  $\glabcms=\sqrt{s}/(2m_p)\simeq 60$ and the rapidity
shift is $\tanh^{-1} \blabcms\simeq 4.8$. With 2.76 TeV lead ions, one has
$\glabcms\simeq 38$ and  $\dylabcms\simeq 4.3$.
In both cases, the central-rapidity region in the  c.m.s., $y_{\rm cms}\simeq 0$, is thus highly boosted at an angle 
w.r.t.  the beam axis of about one degree in the laboratory frame. One can easily access the entire backward 
c.m.s. hemisphere, $y_{\rm cms}<0$, with standard experimental techniques. The forward hemisphere is probably less 
conveniently accessible; the reduced distance from the (extracted) beam axis requires the use of highly segmented 
detectors to deal with the large particle density. We thus consider that 
one can access the region $-4.8\leq y_{\rm cms}\leq 1$ without specific difficulties.
 Such an acceptance covers the bulk of most yields and offers the opportunity of high precision 
measurements in the whole backward hemisphere, down to 
$x_F\to -1$ for multiple systems. For instance, by studying  $\Upsilon$ production at
rapidities of the order of -2.4 in the c.m.s., one can access $x_F$ above $\frac{10}{115} e^{2.4}\simeq 0.95$.

The gluon and heavy-quark distributions in the proton, the neutron and nuclei (see \eg~\cite{Diakonov:2012vb})  could then 
be extracted at mid and large momentum fractions, $x$, by accessing the target-rapidity region. 
We also note that, in principle, in the nuclear case, the physics at $x$ larger than unity 
--which necessarily probe nuclear correlation-- can be accessed. One could study the scale dependence
of nuclear effects in the EMC and Fermi motion region, $0.3 < x < 1$; this may be fundamental to understand  
the connexion between the EMC effect and the importance of short-range correlations.
 
\begin{figure*}[!hbt]
\includegraphics[width=\textwidth]{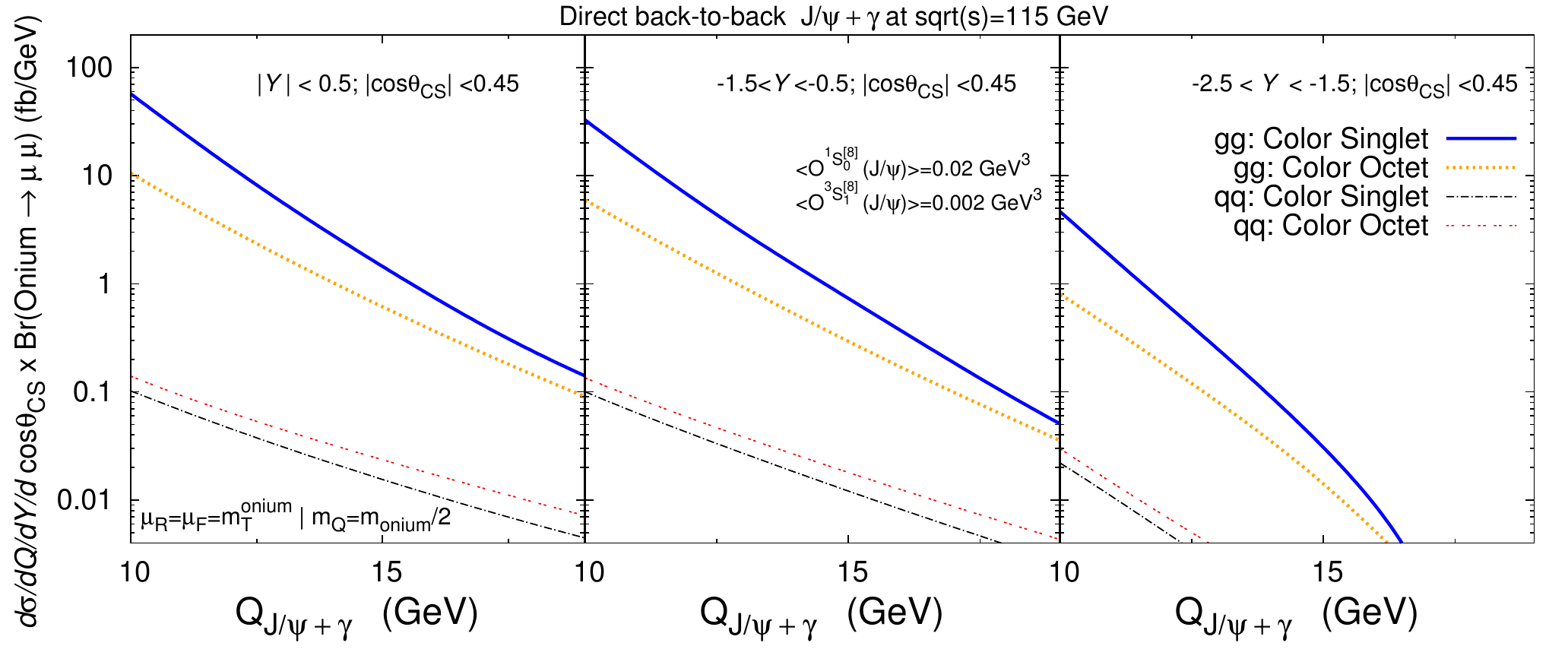}
\caption{Different contributions to the production of an isolated photon back-to-back with a $J/\psi$  
from $g-g$ and $q-\bar q$ fusion  from the CS and CO channels as a function of the invariant mass of the pair $Q_{(J/\psi+\gamma)}$ for
three different rapidity regions (from left to right: $|Y|<0.5$, $-1.5< Y < -0.5$ and $-2.5< Y < -1.5$).}
\label{fig:yield-onium-gamma}
\end{figure*}

In proton-deuteron collisions, unique information on the  momentum distribution of the gluons in the neutron
can be also obtained with quarkonium measurements along the lines of 
E866 for $\Upsilon$~\cite{Zhu:2007mja}. 

More generally, thanks to its high luminosity, AFTER@LHC offers many other opportunities related to 
heavy-flavor production, such as quarkonium-associated production (see \eg~\cite{Lansberg:2014ora,Lansberg:2013qka})
or double-charm baryon production \cite{Chen:2014hqa}. For instance, a very backward measurement of 
$J/\psi+D$ production~\cite{Brodsky:2009cf} could tell us much on the charm quark distribution at large $x$ which is the object of
a longstanding debate~\cite{Pumplin:2005yf,Dulat:2013hea,Hobbs:2013bia,Jimenez-Delgado:2014zga}.

Finally, let us stress that the large number of quarkonia
(approximately $10^9$ $J/\psi$ and $10^6$ $\Upsilon$ per unit of rapidity per 20 fb$^{-1}$) 
to be studied with AFTER@LHC offers the possibility to perform high precision (3-dimensional) measurements of their
polarisation~\cite{Lansberg:2012kf}, which is still the subject of intense debates~\cite{Faccioli:2012nv}. 
In the charmonium case, it is very important to perform measurements on the excited-state polarisation 
($\psi'$, $\chi_c$) to avoid to deal with polarisation-dilution effects from their feed-down which preclude one 
to draw strong conclusions from the RHIC data for instance~\cite{Lansberg:2010vq}.

\section{Spin studies with unpolarised protons: Boer-Mulders functions and related distributions}

\vspace*{-.2cm}
\subsection{Low-{\boldmath $P_T$} quarkonium production}

It has been emphasised in~\cite{Boer:2012bt} that the study of quarkonium 
production at low $P_T$ ($P_T \leq M_{\cal Q}$) can provide information on gluon TMDs, \ie~on $f_1^g(x,k_T,\mu)$ 
and  $h_1^{\perp g}(x,k_T,\mu)$, owing to the simplicity of the LO production mechanism (see \cf{fig:graph-etaQ}). 
Unfortunately, the sole study of $\eta_Q$ would not provide enough
information to determine these TMDs separately.

Subsequently, the study of $\eta_Q$ has successfully been  carried out in the TMD factorisation 
at one loop (NLO) in $\alpha_s$~\cite{Ma:2012hh} giving confidence that such a factorisation does 
hold for these particles. On the contrary, it seems not to be the case for $P$-wave 
production~\cite{Ma:2014oha}.

With $P_T$-integrated cross sections for $\eta_c$ at $y=0$ as large as 1 nb, such studies 
can be envisioned. It is however not clear yet down to which $P_T$ they could be carried out. 
This would depend much on the decay channel, which can be $KK\pi$, $p\bar p$, $\gamma \gamma$, ..., and 
on the detection technique used. Forthcoming simulations will be extremely insightful on this matter.
The very first measurement of inclusive $\eta_c$ hadroproduction by the LHCb 
collaboration~\cite{Aaij:2014bga} gives us confidence that such a measurement is nowadays possible at least in
the $p \bar p$ decay channel.

\begin{figure}[!hbt]
\subfloat[]{\includegraphics[width=2.75cm]{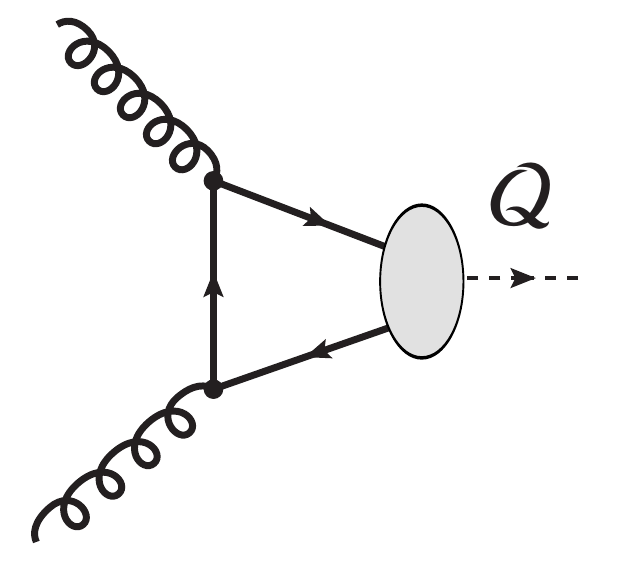}\label{fig:graph-etaQ}}
\subfloat[]{\includegraphics[width=2.75cm]{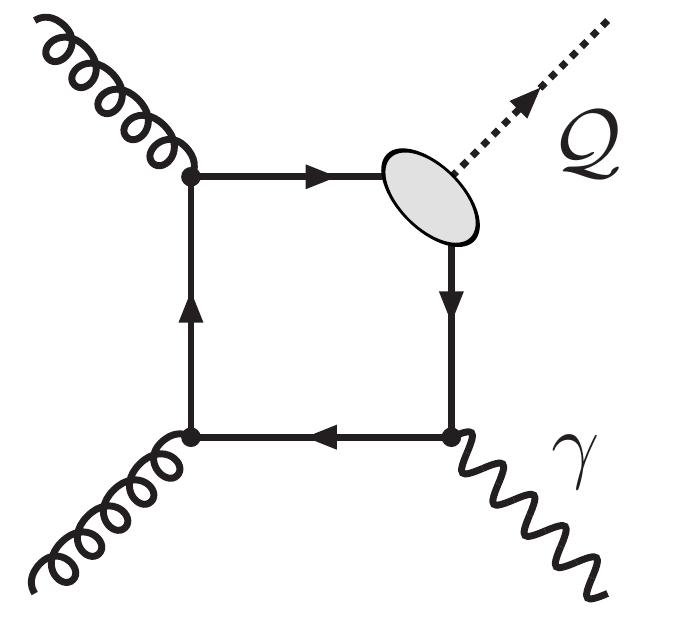}\label{fig:graph-psi-gamma}}
\subfloat[]{\includegraphics[width=2.75cm]{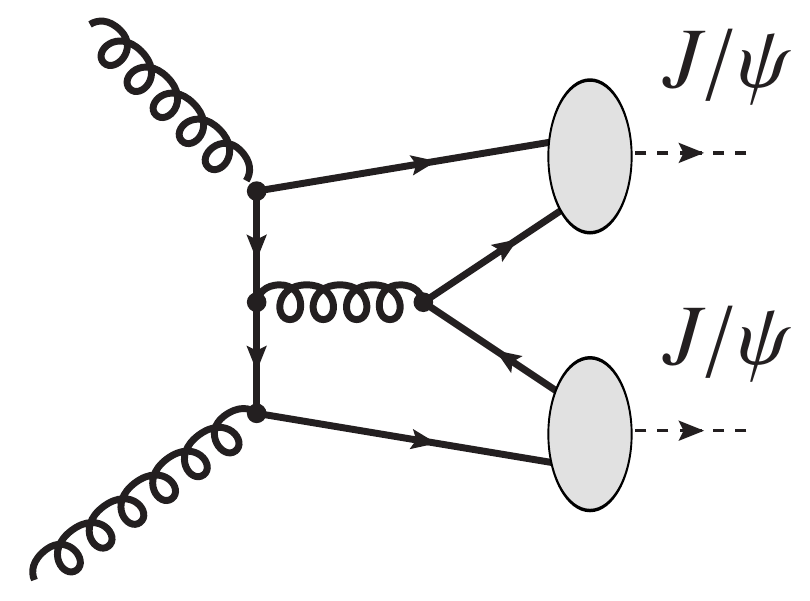}\label{fig:graph-psi-psi}}
\caption{Typical LO Feynman graphs for  (a) $\eta_Q$, (b) $\psi+\gamma$ and (c) $J/\psi$-pair production.}
\label{fig:graphs}
\end{figure}

\vspace*{-.2cm}
\subsection{Back-to-back quarkonium\,{\boldmath $+ \gamma$} production} 

In~\cite{Dunnen:2014eta}, we discussed the possibility of extracting the polarised 
and unpolarised gluon TMDs, $f_1^g(x,k_T,\mu)$ and  $h_1^{\perp g}(x,k_T,\mu)$ 
through the production of a quarkonium associated
with a back-to-back and isolated photon at the LHC, thus with unpolarised protons. Contrary to the inclusive production of 
quarkonium + photon pair, which has been shown to help to disentangle color-octet (CO) from 
color-singlet (CS) contributions~\cite{Li:2008ym,Lansberg:2009db,Li:2014ava}, back-to-back production
tends to be dominated by CS contributions~\cite{Mathews:1999ye} at least at low transverse momenta. In addition, 
it  is less prone to QCD corrections, which simplifies its theoretical study. In the case of 
$J/\psi$ and $\Upsilon$ production, the requirement 
for back-to-back production essentially selects the topologies of graphs \cf{fig:graph-psi-gamma}; these
are dominated by gluon fusion as expected at high energies for heavy-quark production.
If CS contributions are indeed dominant, TMD factorisation should apply and the measurements of the yield
as a function of the transverse-momentum imbalance of the pair, $q_T$, should give, for the first time, a direct 
access to $f_1^g(x,k_T,\mu)$. Moreover, the study of azimuthally-modulated  
moments of the yields as function of $q_T$ should allow one to look for a nonzero linear polarisation of gluons
inside unpolarised protons by extracting $h_1^{\perp g}(x,k_T,\mu)$. 

At the LHC, the study of isolated photons usually imposes
to require a minimal transverse momentum, of the order of 10 GeV. At lower energies, 
it is likely possible to cope with a lower threshold, for instance 4 GeV, 
thanks to the lower particle multiplicities, especially 
in the backward region accessible with AFTER@LHC. It is nonetheless legitimate to wonder whether 
such an observable would still be sensitive to gluons in the region of rather large $x$ values.
Let us recall that $x_{1,2}=Q/\sqrt{s} e^{\pm Y}$. Thus, for $Q=10$~GeV and $Y=-2$, $x_2\simeq 0.65$. 
We have checked the gluon-fusion dominance as illustrated by~\cf{fig:yield-onium-gamma}. 
From left to right, the rapidity $Y$ of the pair is getting more
negative. First, the $q\bar q$ contribution remains negligible --at most a percent of that from $gg$. 
Second, the CS contribution (solid blue curve) remains above the CO one 
(orange dashed curve) up to $Q\simeq 20$~GeV. At $Q \simeq 10$~GeV, the expected CO contribution is less than 
a quarter of the CS one and can be disregarded for a first TMD extraction. In fact, it can also 
simply be removed by isolating the $J/\psi$ as well; this would be required if the measured value of the $q_T$ integrated 
yield was higher than expected and thus indicative of a larger CO yield. Finally, we wish to stress 
that we do not expect higher-twist contributions such as intrinsic-charm (IC) quark coalescence~\cite{Brodsky:1989ex} 
to contribute to (back-to-back) $J/\psi+\gamma$ production.
As opposed to inclusive $J/\psi$ production for which one has $J/\psi+g$ at LO and for which the color of the
non-perturbative IC fluctuations can be bleached by the final-state gluon emission, the emission of
the final-state photon is irrelevant; an additional gluon emission is needed for this mechanism to contribute.

In terms of expected counts, since the differential cross section is on the order of tens of fb/GeV, 
one can reasonably expects a couple of thousands of events per year (\ie~per 20 fb$^{-1}$) 
with a hydrogen target. This is definitely sufficient to look at the $k_T$ dependence 
of $f_1^g(x,k_T,\mu)$ as well as to look at the magnitude of $h_1^{\perp g}(x,k_T,\mu)$ at large $x$.

Another observable where the TMD factorisation should be applicable is $J/\psi$-pair 
production. As for $J/\psi+\gamma$ production, the final state (see \eg~\cite{Lansberg:2013qka})
can be fully color singlet (see \cf{fig:graph-psi-psi}).
Its analysis should also be very well accessible with AFTER@LHC.

\section{Spin studies with polarised protons: Single Transverse Spin Asymmetries}

\subsection{Looking for the gluon Sivers effect and beyond}

STSA were computed for $\eta_Q$ production in the collinear twist-3 approach~\cite{Schafer:2013wca}. In this case, 
the STSA arises from twist-3 quark gluon correlators $T_F(x,x)$, also known as the Efremov-Teryaev-Qiu-Sterman correlators. 
In the TMD factorisation, it is due to the well-know Sivers function. The study of STSA in $\eta_Q$ production is 
particularly interesting because of the possible differences -- as the sign mismatch first discussed in~\cite{Kang:2011hk} 
-- between both these approaches, 
since the Collins effect is not expected to contribute here. Such a measurement is 
certainly possible with AFTER@LHC with a transversely polarised target. An important point 
is to be able to carry out such a measurement down to low $P_T$ where both approaches are applicable and
can legitimately be compared. In any case, such a measurement would be extremely useful to
tell whether or not such a gluon Sivers effect does exist.

The PHENIX collaboration~\cite{Adare:2010bd} has measured the STSA in $J/\psi$ production at $\sqrt{s}=200$~GeV. 
They reported a value of $A_N$ compatible with 0 with a slightly negative central value. More precise data 
are definitely needed. AFTER@LHC can certainly push far forward the precision limit on such a measurement.
It is noteworthy to emphasise the possibility to collide lead ions on a polarised target since a number 
of theoretical ideas have been proposed lately in the case of $p^\uparrow A$ collisions.

Further measurements can be carried out with a polarised target. By measuring 
the angular correlations in $\psi+\gamma$ production involving $\phi$, 
the azimuthal Collins-Soper angle and $\phi_{ST}^{qT}$, 
the angle between the $q_T$ of the pair and the transverse polarisation vector of the proton,
one gain access to $f_{1T}^{\perp g}$, $h^g_{1T}$ and
$h_{1T}^{\perp g}$, in addition to $f^g_1$ and $h_1^{\perp g}$ which are accessible without target polarisation.

\subsection{Quark Sivers effect}

AFTER@LHC is also a good playground to study the quark Sivers effect by measuring STSA in Drell-Yan pair 
production~\cite{Liu:2012vn}. Such studies would nicely complement the forthcoming DY STSA measurements 
in pion-induced reaction at COMPASS~\cite{Quintans:2011zz} and two proposals at Fermilab, 
P1027~\cite{Isenhower:2012vh} with a polarised beam to study the large $x^\uparrow$ domain
and  P1039~\cite{Brown:2014sea} with a polarised target for lower $x^\uparrow$. 

As for now, the main objectives of such measurements is to verify the {\it prima facie} robust prediction of QCD according 
which the Sivers function changes sign, when going from semi-inclusive DIS to DY pair production~\cite{Collins:2002kn}.

\section{Conclusion}

In conclusion, a fixed-target experiment using the LHC beams can provide us with extremely complementary measurements
to those made at RHIC and at lower energy fixed target projects, such as COMPASS and P1027 or P1039, which are dedicated to 
spin physics or TMD extraction. 

At high energies, the fixed-target mode is very well adapted for measurements at large $x$ in the target. 
The latter can be polarised and this opens the path to the study of target-spin asymmetries at large 
$x^\uparrow$, where they are expected to be the largest. Moreover, as we stressed, a number of spin-related measurements
can also be carried without a target polarisation, by taking advantage of the high luminosities 
and a low-$P_T$ acceptance and by looking
at transverse-momentum dependent phenomena, encapsulated in the TMDs.

{\small {\bf Acknowledgements.}
This research [SLAC-PUB-16099] was supported in part by the French CNRS via the grants PICS-06149 Torino-IPNO, 
FCPPL-Quarkonium4AFTER \& PEPS4AFTER2, by the Tournesol 2014 Wallonia-Brussels-France Cooperation Programme, 
and by the Department of Energy, contract DE-AC02-76SF00515. 
}
\vspace*{-.2cm}

%
% BibTeX or Biber users please use (the style is already called in the class, ensure that the "woc.bst" style is in your local directory)
% \bibliography{name or your bibliography database}
%
% Non-BibTeX users please use
%

\end{document}